\documentclass[12pt,a4paper]{article}
\input epsf
\usepackage{amsmath,amsfonts,epsfig,color,latexsym}
\usepackage{amssymb}
\usepackage[english, USenglish]{babel}
\usepackage{epsfig}
\usepackage{a4wide}
\usepackage{rotating}
\usepackage{slashed}
\newcommand{\N}{\mathcal{N}}

\csname @addtoreset\endcsname{equation}{section}
\author{\\[-0.3cm]\Large I.~Antoniadis\footnote{{\tt ignatios.antoniadis@cern.ch}. On leave from CPHT (UMR CNRS 7644) Ecole Polytechnique, F-91128 Palaiseau.}~, S.~Hohenegger\footnote{{\tt stefanh@itp.phys.ethz.ch}}~, K.S.~Narain\footnote{{\tt narain@ictp.trieste.it}}~, T.R.~Taylor\footnote{{\tt taylor@neu.edu}. On sabbatical leave from Department of Physics,
Northeastern University, Boston, MA 02115,  U.S.A.}\\
}
\date{}
\graphicspath{{figures/}}
\begin{document}
\begin{titlepage}
\title{\vspace{-1cm}\begin{flushright}{\small{\vspace{-0.8cm} CERN-PH-TH/2010-061\\\vspace{-0.3cm}
LMU-ASC 09/10}}\end{flushright}
\vspace{0.8cm}\bf{Deformed Topological Partition Function and Nekrasov Backgrounds}}
\maketitle
\begin{center}
\renewcommand{\thefootnote}{\fnsymbol{footnote}}\vspace{-0.5cm}
\footnotemark[1]Department of Physics, CERN - Theory Division, CH-1211 Geneva 23, Switzerland\\[0.5cm]
\vspace{-0.3cm}
\footnotemark[2]Institut f\"ur Theoretische Physik, ETH
  Z\"urich,
CH-8093 Z\"urich, Switzerland\\[0.5cm]
\vspace{-0.3cm}
\footnotemark[3]High Energy Section, The Abdus Salam International Center for Theoretical Physics,
\\Strada Costiera, 11-34014 Trieste, Italy\\[0.5cm]
\vspace{-0.3cm}\footnotemark[4]Center for Advanced Studies, Ludwig-Maximilians-Universit{\"{a}}t,\\
Seestra{\ss}e 13, 80802 Munich, Germany\\[0.5cm]
\end{center}
\begin{abstract}\noindent
A deformation of the $\N=2$ topological string partition function is
analyzed by considering higher dimensional F-terms of the type
$W^{2g}\Upsilon^n$, where $W$ is the chiral Weyl superfield and each
$\Upsilon$ factor stands for the chiral projection of a real function
of $\N=2$ vector multiplets. These terms generate physical amplitudes 
involving two anti-self-dual Riemann tensors, $2g-2$ anti-self-dual 
graviphoton field strengths and $2n$ self-dual field strengths from 
the matter vector multiplets. Their coefficients $F_{g,n}$  generalizing the genus 
$g$ partition function $F_{g,0}$ of the topological twisted type II theory,
can be used to define a generating functional by introducing
deformation parameters besides the string coupling. Choosing all matter field strengths to be that of the dual heterotic
dilaton supermultiplet, one obtains 
two parameters that we argue should correspond to the deformation
parameters of the Nekrasov partition function in the field theory
limit, around the conifold singularity. Its perturbative
part can be obtained from the one loop analysis on the heterotic side. This has been computed in
\cite{Morales:1996bp} and in the field theory limit shown to be given by the radius deformation of
$c=1$ CFT coupled to two-dimensional gravity. Quite remarkably this result reproduces
the gauge theory answer up to a phase difference that may be
attributed to the regularization procedure. The type II results 
are expected to be exact and should also capture the part that is  non-perturbative 
in heterotic dilaton.
\end{abstract}
\thispagestyle{empty}
\end{titlepage}

\noindent
\section{Introduction}
$\N=2$ supersymmetric theories provide an interesting arena for
studying non perturbative dynamics of field and string theories (in
the absence or presence of gravity), exhibiting a remarkable web of
dualities among very different systems. It was initiated with the
computation of the prepotential that describes the exact two
derivative effective action in the Coulomb phase of $\N=2$ super Yang
Mills theory~\cite{sw}. This result was subsequently generalized in the
presence of gravity~\cite{kv}, using string duality between heterotic 
compactifications on $K3\times T^2$, that describe the perturbative
expansion around an enhanced symmetric point of the gauge symmetry, 
and type II on a Calabi-Yau threefold which is an appropriate $K3$ 
fibration, that provides the exact answer around the conifold
singularity where massless charged hypermultiplets appear as non
perturbative states~\cite{stro}. The field theory limit corresponds to
the double scaling where the string length and the distance from the 
conifold point go to zero, keeping fixed the dynamical Yang Mills scale.

On the other hand, it was known that the prepotential is the first of
a series of $\N=2$ F-terms of the type $F_gW^{2g}$ where $W$ is the 
chiral Weyl superfield, containing the graviphoton anti-self-dual 
field strength as the lowest component, and $F_g$ are analytic
functions of the vector moduli~\cite{Antoniadis:1993ze,bcov}; they are 
computed by the genus $g$ partition function of the twisted Calabi-Yau 
sigma model and have a non analytic part due to a holomorphic anomaly 
that satisfies an equation which provides interesting recursion
relations. Moreover, their behavior near the conifold singularity is 
governed by $c=1$ string theory at the self-dual radius~\cite{c1,gv}. 
This was explicitly checked on the heterotic side, where $F_g$'s start 
receiving contributions already at one loop~\cite{Antoniadis:1995zn}. 
It was indeed shown that near the enhanced symmetry point the leading 
singularity of $F_g$ (for $g>1$) is a universal pole of order $2g-2$ 
with a coefficient given by the Euler number of the genus-$g$ Riemann surface.

Another important result was the explicit computation of the $\N=2$
gauge theory partition function~\cite{Nekrasov:2002qd} by performing the sum over all
instanton contributions. This was achieved by 
calculating the instanton partition function as a function of two complex parameters
$\epsilon_{1,2}$ that regulate the integral over the instanton moduli space,
and interpreting it as a six-dimensional theory in the so-called $\Omega$
supergravity background~\cite{epsilons,LNS}.
It was then shown that upon extracting a
divergent volume factor $1/\epsilon_1\epsilon_2$, one can reproduce
the Seiberg-Witten prepotential in the limit $\epsilon_{1,2}\to
0$. Moreover, it was argued that all higher $F_g$'s can be reproduced
in the field theory limit, by identifying $\epsilon_1=-\epsilon_2$
with the topological string coupling. 

The perturbative part of the gauge theory for $\epsilon_+ \equiv
(\epsilon_1+\epsilon_2)/2=0$ was exactly the same as the leading
terms near the $SU(2)$ enhancement in
the one loop heterotic result for the generating function
$\sum_g \epsilon_-^{2g} F_g$ \cite{Antoniadis:1995zn}. In fact the latter 
was just given by the Schwinger like formula for a one loop diagram for 
states that become massive in the Coulomb branch of ${\cal N}=2$ $SU(2)$ gauge
theory,  coupled to a constant anti-self dual background
field strength. 
In the Gopakumar-Vafa
reformulation \cite{Gopakumar:1998ii,Gopakumar:1998jq} of the topological partition function the
generating function of the A-model $F_g$  is reproduced by integrating all
the massive BPS states which are the  
D-brane states wrapping on 2-cycles of the Calabi-Yau space in the
background of a constant anti-self dual graviphoton field
strength; the contribution of each of these states being given by  the
Schwinger formula. 
Due to anti-self duality, the graviphoton field strength couples to the spin of
the D-brane states in one of the $SU(2)$ in the $SO(4)$ Lorentz group.
Using the technique of the topological vertex for toric non-compact
Calabi-Yau spaces that can be used to geometrically engineer gauge
theories, it was shown in \cite{Hollowood:2003cv}
that $F_g$ captures 
also the non-perturbative part of the gauge theory for
$\epsilon_1=-\epsilon_2$ where the combination $\epsilon_1-\epsilon_2$
has the interpretation of the background of the anti-self dual graviphoton
field strength. This suggests that $\epsilon$'s
are not just regularization parameters and opens the question what is
the string theory interpretation of the second parameter
$\epsilon_+$. 

In \cite{Iqbal:2007ii}, a generalization of the Gopakumar-Vafa formula was proposed which includes 
coupling to spins of both $SU(2)$ of the D-brane states. The
couplings correspond to backgrounds involving both self-dual and
anti-self dual gauge field strengths. Using the formulation of a
refined topological vertex for non compact toric Calabi-Yau spaces   
it was shown that this generalized formula reproduces the
non-perturbative
part of the gauge theory result for arbitrary $\epsilon_-$ and
$\epsilon_+$. This raises the question what string theory amplitude
gives rise to this generalized Gopakumar-Vafa formula and which
string effective action term it computes.

In this work, we address the above question by studying possible
deformations of the topological string and we make a proposal based on 
some plausible arguments. Indeed, the $\Omega$-background for the 
six-dimensional metric corresponds to constant field strengths for 
some Kaluza-Klein gauge fields that include the $\N=2$ graviphoton, 
as well as another direction, such that part of the supersymmetry 
is preserved. This leads us to consider $\N=2$ higher dimensional 
F-terms that generalize the series $F_gW^{2g}$ and involve only vector 
multiplets. By analogy with a similar study which has been performed 
in the past for $\N=1$ amplitudes~\cite{Antoniadis:1996qg}, we
consider a `dressing' of $W^{2g}$ with chiral projections of functions 
of vector multiplets that we call generically $\Upsilon$:
$W^{2g}\Upsilon^n$. These terms generate, in particular, amplitudes 
of the form $F_{g,n}(R^-)^2(T^-)^{2g-2}(F^+)^{2n}$, where $R$ denotes the
Riemann tensor, $T$ the graviphoton, $F$ the remaining matter gauge 
field strengths, while the superscripts $-$ and $+$ denote their 
anti-self-dual and self-dual projections, respectively. We have here
suppressed the anti-holomorphic vector moduli indices both in the
matter field strengths $F$ as well as in the coefficient $F_{g,n}$
which depend on both holomorphic and anti-holomorphic vector moduli.\footnote{The dependence of $F_{g,n}$ on the anti-holomorphic vector moduli is controlled by certain differential
equations that  arise from their origin in the F-term~\cite{Antoniadis:1996qg}. We leave a
discussion of these classical equations as well as the possible
anomalies at the string level for future.}  On the type II 
side, these amplitudes are topological, given essentially by the
square of the left-moving (supersymmetric) part of the corresponding 
heterotic higher dimensional terms~\cite{Antoniadis:1996qg}.

Motivated again by the $\Omega$-background, we are led to consider the 
case where all matter gauge field strengths are those of the heterotic 
dilaton vector supermultiplet, arising from compactification of the
$\N=1$ six-dimensional theory on $T^2$. Such amplitudes have been
considered before in~\cite{Morales:1996bp} and were studied on the 
heterotic side. Their behavior near the enhanced symmetric point was 
found to be determined by the radius deformation of the $c=1$ string 
theory. Here, we compare their expression with the perturbative limit 
of the Nekrasov partition function and we find a remarkable agreement 
up to a phase factor that could be attributed to a difference in the 
field theory regularization. More precisely, the string theory
amplitudes in the heterotic perturbative limit are invariant under 
parity which changes the sign of one of the two $\epsilon_1$ or $\epsilon_2$ but 
leaves the other invariant. On the other hand, the field theory 
result breaks the symmetry due to the additional phase, although 
the corresponding action is invariant~\cite{Nekrasov:2003rj}, up 
to the $\theta$-term which is relevant only for non-perturbative
contributions.

The plan of this paper is the following. In
Section~\ref{Sect:StringAmplitude}, we define the new amplitudes and 
describe the effective Lagrangian. Then, we compute the corresponding 
amplitudes in string theory  on the type II side
(Section~\ref{Sect:TypeII}) and review the corresponding heterotic
computation of ~\cite{Morales:1996bp} in
Section~\ref{Sect:HetAmp}. In type~II theory, they are topological 
and are expected to be exact at genus $g$, while on the heterotic side they 
start receiving contributions at one loop. Finally, we take the field 
theory limit near the enhanced symmetric point and compare the
resulting expression with the perturbative part of Nekrasov's partition function.

\section{Higher Derivative Couplings in the String Effective Action}\label{Sect:StringAmplitude}
In this section we discuss a class of amplitudes in $\N=2$
compactifications of string theory, which we will compare to the
Nekrasov-partition function as discussed in
\cite{Nekrasov:2002qd,epsilons,LNS,Nekrasov:2003rj}. 
In \cite{Antoniadis:1993ze} a particular class of higher derivative chiral F-terms
\begin{align}
\mathcal{I}_g=\int d^4x\int d^4\theta F_g(X) W^{2g}\,,\label{ChiralFterm}
\end{align}
in the effective $\N=2$ Poincar\'e supergravity action has been
considered. Here $W$ is the Weyl superfield which has Weyl weight
$w=1$. It has the following expansion (we only display the bosonic components)
\begin{align}
W^{ij}_{\mu\nu}=\epsilon^{ij}T_{\mu\nu}^- -R_{\mu\nu\lambda\rho}^-
(\theta^i\sigma^{\lambda\rho}\theta^j)
+\ldots,
\end{align}
where $T_{\mu\nu}^-$ is the anti-self-dual graviphoton field
strength and $R_{\mu\nu\lambda\rho}^-$ is the anti-self-dual Riemann
tensor. In this respect $\mu,\nu$ denote four-dimensional Lorentz
indices, while $i,j\in SU(2)_R$ are indices of the $\N=2$ R-symmetry group.

Furthermore, $F_g$ is an analytic homogeneous function of degree
$(2-2g)$ of the ``reduced'' $\N=2$ chiral superfields $X^I$
\begin{align}
X^I=\phi^I+\frac{1}{2}F^{-I}_{\lambda\rho}\epsilon_{ij}
(\theta^i\sigma^{\lambda\rho}\theta^j)+\ldots\,,
\end{align}
which have Weyl weight $w=1$ and chiral weight $c=-1$. The field
$X^{0}$ acts as a compensator and the unconstrained physical scalars
(the moduli) are parameterized by $\hat{\phi}^I=\phi^I/\phi^0$ for
$I=1,\ldots,h$. $F^{-I}_{\lambda\rho}$ are the corresponding anti-self-dual gauge
field strengths.
Notice in particular that the superfields
\begin{align}
\hat{X}^I=\frac{X^I}{X^0}\,,
\end{align}
(as well as arbitrary functions thereof) have vanishing Weyl  and
conformal weight. The functions $F_g$ have been shown to be computed
by the partition function of the $\N=2$ topological string.
Furthermore, in the field theory limit i.e. decoupling the
gravitational sector apart from their backgrounds, the generating
function $F(\lambda)=\sum_g \lambda^{2g} F_g$ is related to the $\N=2$
gauge theory partition function in the presence of anti-self-dual graviphoton
field strength ($\epsilon_1=-\epsilon_2$ in the notation of
\cite{Nekrasov:2002qd,Nekrasov:2003rj}).

In this note we will consider a slight generalization of
(\ref{ChiralFterm}). To this end, we introduce the superconformal
chiral projection operator (see e.g.~\cite{deRoo:1980mm})
\begin{align}
&\Pi=(\epsilon_{ij}\bar D^i\bar\sigma_{\mu\nu}\bar D^j)^2\,,
&&\text{with} &&\Pi (X^I)=0~~\text{and}~~\Pi (X^I)^\dagger=96\Box X^I\,.
\end{align}
From this relation we can read off that $\Pi$ has Weyl weight $w=2$
and conformal weight $c=-2$. Therefore, for an arbitrary function
$G(\hat{X}^I,(\hat{X}^I)^\dagger)$ the superfield
\begin{align}
\Upsilon=\Pi \frac{G(\hat{X}^I,(\hat{X}^I)^\dagger)}{(X^0)^2}\,,
\end{align}
will have vanishing Weyl  and conformal weights. Notice in particular
that the first term in the expansion of $\Upsilon$ will be given by
\begin{align}
\Upsilon=\frac{\partial^2 G(\hat{\phi},\hat{\phi}^\dagger)}{\partial
(\hat{\phi}^I)^{\dagger}\partial (\hat{\phi}^J)^{\dagger}}\,(F^{+I}
F^{+J})+\ldots\,.
\label{upsilonexp}
\end{align}
We can therefore generalize (\ref{ChiralFterm}) in the following way
\begin{align}
\mathcal{I}_{g,n}=\int d^4x\int d^4\theta ~  W^{2g} \Upsilon^n
\sim\int d^4x F_{g,n}(\hat{\phi}, \hat{\bar{\phi}}) (R^-)^2 (T^-)^{2g-2}(F^+)^{2n}+\ldots\,.
\label{ChiralFtermPi}
\end{align}
Here we have suppressed the $2n$ anti-holomorphic vector indices in
$F_{g,n}$ that contract the corresponding indices on the $2n$ matter
gauge fields and the dots denote further terms, which will play no role in this
note. The anti-holomorphic moduli dependence in $F_{g,n}$ comes from
$\Upsilon$ as shown in (\ref{upsilonexp}).

We would like to point out that terms involving both anti-self-dual and
self-dual graviphoton field strengths cannot be written as a chiral
F-term. This is because the self-dual graviphoton field strength is
the lowest component of the anti-chiral Weyl multiplet. Thus $n$
chiral projectors acting on functions of anti-chiral Weyl superfields
will pick out $2n$ self dual Riemann tensors.

\section{The Amplitude in Type II String Theory}\label{Sect:TypeII}
We will now try to recover the higher derivative terms
(\ref{ChiralFtermPi}) in string theory, starting
with type II compactified on an arbitrary Calabi-Yau manifold. The
couplings (\ref{ChiralFterm}) have been recovered as $g$-loop
amplitudes in \cite{Antoniadis:1993ze}. Since (by construction) the
superfields $\Upsilon$ which we have added in (\ref{ChiralFtermPi})
have no Weyl weight, we are led to conclude that also $F_{g,n}$
corresponds to a $g$-loop amplitude in the type II theory. We will
now try to explicitly verify this conjecture.

\begin{figure}[ht]
\begin{center}
\input{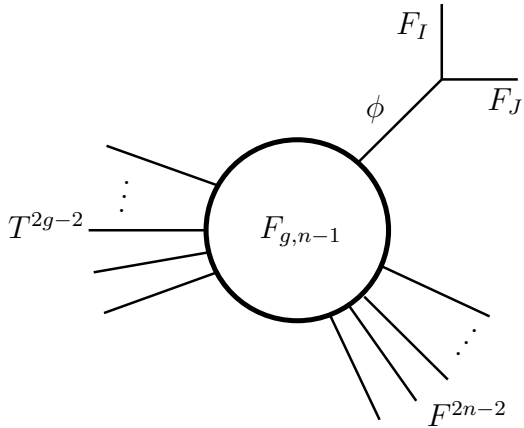}
\caption{Schematic drawing of the reducible diagram: The gauge kinetic term $F_I.F_J$ is connected to $F_{g,n-1}$ via the exchange of the scalar $\phi$.}
\label{Fig:Diagram}
\end{center}
\end{figure}

General expressions for the vertex operators which are needed to
compute the amplitude corresponding to the component term that we
have explicitly displayed in (\ref{ChiralFtermPi}) can for example
be found in \cite{Antoniadis:1993ze}. Taking into account that the
gauge fields are self-dual in (\ref{ChiralFtermPi}), while the
Riemann tensors and graviphotons are anti-self-dual, we are
led to specifically consider the following amplitude
\begin{align}
F_{g,n}=\langle V_{(R)}(p_1)V_{(R)}(\bar{p}_1)\prod_{i=1}^{g-1}
V^{(-1/2)}_{(T)}(\bar{p}_1^i)V^{(-1/2)}_{(T)}(\bar{p}_1^i)
\prod_{j=1}^nV^{(-1/2)}_{(F),I_j}(p_1^{j})V^{(-1/2)}_{(F),J_j}
(\bar{p}_1^{j})\rangle_{\text{g-loop}}^{\text{type II}}\,,\label{TypeIIAmplitude}
\end{align}
where we have picked the following helicity and momentum
configurations in the various vertices
\begin{center}
\begin{tabular}{ll}
gravitons: & $V_{(R)}(p_1)=:\left(\partial Z^2-ip_1\chi^1\chi^2\right)
\left(\bar{\partial} Z^2-ip_1\tilde{\chi}^1\tilde{\chi}^2\right)
e^{ip_1Z^1}(z_1,\bar{z}_1):\,,$\\
& $V_{(R)}(\bar{p}_1)=:\left(\partial \bar{Z}^2-i\bar{p}_1\bar{\chi}^1
\bar{\chi}^2\right)\left(\bar{\partial} \bar{Z}^2-i\bar{p}_1
\bar{\tilde{\chi}}^1\bar{\tilde{\chi}}^2\right)
e^{i\bar{p}_2\bar{Z}^2}(z_2,\bar{z}_2):\,,$\\
&\\
graviphotons: & $V^{(-1/2)}_{(T)}(p_1)=:p_1
e^{-\frac{1}{2}(\varphi+\tilde{\varphi})}S_1\tilde{S}_1\Sigma(x,\bar{x})
 e^{ip_1 Z^1}:\,,$\\
&
$V^{(-1/2)}_{(T)}(\bar{p}_1)=:\bar{p}_1e^{-\frac{1}{2}(\varphi+\tilde{\varphi})}
S_2\tilde{S}_2\Sigma(y,\bar{y}) e^{i\bar{p}_1\bar{Z}^1}:\,,$\\
&\\
gauge fields: &
$V^{(-1/2)}_{(F),I}(p_1)=:p_1e^{-\frac{1}{2}(\varphi+\tilde{\varphi})}
S^{\dot{1}}\tilde{S}^{\dot{1}}\Sigma_{ I}(u,\bar{u}) e^{ip_1Z^1}:\,,$\\
& $V^{(-1/2)}_{(F),J}(\bar{p}_1)=:c\tilde{c}\bar{p}_1
e^{-\frac{1}{2}(\varphi+\tilde{\varphi})}S^{\dot{2}}\tilde{S}^{\dot{2}}
\Sigma_{ J}(v,\bar{v}) e^{i\bar{p}_1\bar{Z}^1}:\,.$
\end{tabular}
\end{center}
${}$\\[10pt]
Here we have adopted the convention to denote right moving degrees of
freedom with a tilde. We have introduced a complex basis for the
space-time (Euclidean) coordinates $(Z^1,\bar{Z}^1,Z^2,\bar{Z}^2)$
as well as their fermionic partners $(\chi^1,\bar{\chi}^1,\chi^2,
\bar{\chi}^2)$. If we bosonize the latter in terms of $\phi_{1,2}$
we can write for the space-time spin fields
\begin{align}
&S_1=e^{\frac{i}{2}(\phi_1+\phi_2)}\,,&&\text{and} &&S_2=
e^{-\frac{i}{2}(\phi_1+\phi_2)}\,,\\
&S^{\dot{1}}=e^{\frac{i}{2}(\phi_1-\phi_2)}\,,&&\text{and}
&&S^{\dot{2}}=e^{-\frac{i}{2}(\phi_1-\phi_2)}\,.
\end{align}
Recall that the self-dual and anti-self-dual tensors transform under
the Euclideanized Lorentz group $SO(4)=SU(2)\times SU(2)$ as $(3,1)$
and $(1,3)$ representations respectively.
For the gauge field vertices above we have chosen the helicity combination
$S^{\dot{1}} \tilde{S}^{\dot{1}}$ and $S^{\dot{2}}
\tilde{S}^{\dot{2}}$
which corresponds to helicity $+1$ and $-1$ of the corresponding
$SU(2)$. We have distinguished the vector indices for these two
helicities by the indices $I_i$ and $J_j$ with $i,j=1,\ldots, n$ respectively.
Generalization to the helicity $0$ would be given by the
vertices $S^{(\dot{1}} \tilde{S}^{\dot{2})}$ and an analysis for this
case can also be carried out. At the end we will point out the
resulting modification. Note that when one of the vertices labelled by
index $I$ approaches another labelled by $J$, they could form a
Lorentz invariant combination and could result in a divergence
associated with the reducible graph connecting the gauge kinetic term
$F_I.F_J$ with the amplitude $F_{g,n-1}$ via the exchange of
a scalar modulus (see Figure~\ref{Fig:Diagram}). We will comment on this further at the end of this section.

$\Sigma$ is a field in the internal $\N=2$ sector with
$U(1)$-charges $(3/2,-3/2)$ in the type IIA theory and $(3/2,3/2)$
in type IIB. Bosonizing the $U(1)$-current $J$ of the internal $\N=2$
SCFT\footnote{The full SCFT will be spanned by the energy momentum
tensor $T$, two spin-3/2 supercurrents $G^\pm$ and the $U(1)$-current
$J$.} by $H$ we can explicitly write
\begin{align}
\Sigma(z,\bar{z})=e^{\frac{i\sqrt{3}}{2}\left(H(z)\mp\tilde{H}(\bar{z})
\right)}\,,
\end{align}
where the upper sign corresponds to IIA and the lower to IIB (for
further details see e.g.~\cite{Antoniadis:1993ze}). The operators
$\Sigma_{ I}$ have charge $(1/2,-1/2)$ [$(1/2,1/2)$] in type IIA
(type IIB) respectively and are given by
\begin{align}
\Sigma_{
  I}(z,\bar{z})=\lim_{w\rightarrow z} |w-z|e^{\frac{\sqrt{3}i}{2}(H(w)\mp\tilde{H}(\bar{w}))}
\Psi_{ I}(z,\bar{z})\,,
\end{align}
where  $\Psi_{ I}$ are (anti chiral, chiral) (or (anti chiral, anti-chiral))
primary operators in IIA (IIB).
Besides this, observe that we have put $c$-ghosts
together with half of the vertices of the $\N=2$ matter vectors.
Because of this, the corresponding positions will not be
integrated over the world-sheet, but instead we will have $n$
additional Beltrami-differentials $\mu$ corresponding to the punctures.

Finally, the graviton vertices are written in the $(0)$-ghost picture
while all remaining vertices are in the $(-1/2)$-picture which
necessitates the insertion of $(g-1+n)$ picture changing operators
(PCO). Taking into account that we need to insert further $(2g-2)$
PCO which arise after integrating out the superghosts, the total
number of PCO will be $3g-3+n$ and they will be inserted at
world-sheet positions $r_{a=1,\ldots,3g-3+n}$.

After these preliminaries we will now extract the piece of the
correlator which is proportional to $p_1^2\bar{p}_1^2\prod_{i=1}^{g-1}
\bar{p}_1^i\bar{p}_1^i\prod_{j=1}^np_1^j\bar{p}_1^j$. We are thus led to consider
\begin{align}
F_{g,n}=\langle&\bigg|\prod_{k=1}^{3g-3+n}(\mu_kb)e^{i(\phi_1+\phi_2)}
(z_1)e^{-i(\phi_1+\phi_2)}(z_2)\prod_{i=1}^{g-1}e^{-\frac{\varphi}{2}}
S_1e^{\frac{i\sqrt{3}}{2}H}(x_i)\prod_{j=1}^{g-1}e^{-\frac{\varphi}{2}}
S_2e^{\frac{i\sqrt{3}}{2}H}(y_i)\bigg|^2\cdot&\nonumber\\
&\cdot\prod_{k=1}^n\bigg|e^{-\frac{\varphi}{2}}S^{\dot{1}}\bigg|^2
\Sigma_{I_k}(u_k)\prod_{l=1}^n \bigg|c
e^{-\frac{\varphi}{2}}S^{\dot{2}}\bigg|^2
\Sigma_{J_l}(v_l)\bigg|\prod_{a=1}^{3g-3+n}e^{\varphi}T_F(r_a)\bigg|^2\rangle\,,
\label{TypeIIIntegral}
\end{align}
where the absolute square takes into account the contribution of the
right-moving side and there are implicit world-sheet integrals over $z_{1,2},x,y,u$. Comparing this expression to
\cite{Antoniadis:1996qg} we can see that the left moving piece is
identical to the supersymmetric contribution of a similar class of $\N=1$
heterotic topological amplitudes. Since left and right movers in
(\ref{TypeIIIntegral}) are essentially identical, we can immediately
state the complete result
\begin{align}
F_{g,n}=\int_{\mathcal{M}_{(g,n)}}\langle\prod_{k=1}^{3g-3+n}|(\mu_k\cdot
G^-)|^2
\prod_{k=1}^{n}\int
\Psi_{I_k}\prod_{l=1}^n \hat{\Psi}_{J_l}\rangle_{\text{top}}\,.
\label{TopCorrelator}
\end{align}
Here again the absolute square takes care of the corresponding
right-moving operators and $\hat{\Psi}_{ J}=\oint dz\rho(z)\oint
d\bar{z} \tilde{\rho}(\bar{z}) \Psi_{ J}$ with
$\rho$ denoting the unique left  moving operator with charge  $+3$ and dimension $0$ in the
twisted internal theory and $\tilde{\rho}$ denoting the right-moving operator
with charge $(-3)$ for IIA and $(+3)$ for IIB. While the operators $\Psi_I$ have
charge $(-1,+1)$ (and $(-1,-1)$) in IIA (IIB) and dimension $(1,1)$ with respect to the left and
right moving $U(1)$ and the twisted Virasoro generators, the
operators $\hat{\Psi}_J$ have charge $(+2,-2)$ (and $(+2,+2)$)  and dimensions
$(0,0)$ in IIA (IIB). Note that $S^{\dot{1}} \tilde{S}^{\dot{1}}\Sigma_{I}$ of the physical
vertex is replaced by $\Psi_I$ in the twisted theory while
$S^{\dot{2}} \tilde{S}^{\dot{2}}\Sigma_{J}$ is replaced by
$\hat{\Psi}_{J}$.\footnote{ Had we started with the helicity $0$ physical vertex operator
$S^{(\dot{1}} \tilde{S}^{\dot{2})}\Sigma_{K}$ then this would have given
rise to the operator $(\oint dz \rho(z)+ \oint d\bar{z}
\tilde{\rho}(\bar{z})) \Psi_K$ in the twisted theory.}

The short distance singularity mentioned
earlier can arise in general when the operator $\Psi_{I}$ approaches
$\hat{\Psi}_J$ if the kinetic term $F_I.F_J$ is non-vanishing. This
singularity is of the form $1/|z|^2$ and is due to the exchange of a
modulus
that couples to $F_I.F_J$ (see Figure~\ref{Fig:Diagram}). These
reducible graphs have to be subtracted in order to compute the
effective action terms $F_{g,n}$. However for some special cases one
can choose the matter gauge fields in such a way that there are no
such reducible graphs:
\begin{enumerate}
\item  When the Calabi-Yau space is an orbifold $(T^2\times T^2\times T^2)/G$ for some
orbifold group $G$ (e.g. $G= \mathbb{Z}_3$). Denote by $F_i$ for $i=1,2,3$ the
gauge fields corresponding to the moduli of the three $T^2$. The
structure constants $C_{ijk}$ are non-zero only if $i,j,k$ are
different from each other. This means that the gauge kinetic term
$F_i.F_j$ is
non-zero only if $i \neq j$. Choosing all the matter gauge fields
coming from the same $T^2$ will then have no short distance
singularities.
\item A more interesting case is  when the type II theory is dual to a
heterotic $(4,0)$ compactification. The vector modulus $S$ dual to
the heterotic
dilaton  appears linearly in the prepotential
up to exponentially suppressed terms $e^{i S}$. Therefore for large Im$
S$ which corresponds to the weak coupling limit in the heterotic
theory, the singularities will disappear.

We should however emphasize
that the type II result for $F_{g,n}$ is expected to be exact. This can for example
be seen from the effective supergravity action (\ref{ChiralFtermPi}):
The type~II dilaton is the lowest component of an $\N=2$
hypermultiplet. Since in standard $\N=2$ superspace the latter cannot
appear in chiral F-terms in a consistent manner, it follows that
$F_{g,n}$ is independent of the type~II dilaton. Therefore it will
only receive contributions at the $g$-loop level. Notice that this
argument will not go through in the dual heterotic theory since
$F_{g,n}$ will indeed have a non-trivial $S$-dependence. Therefore the
dual heterotic amplitudes will receive also non-perturbative
contributions. However, in order to extract these from the exact type
II result one would have to subtract the reducible terms that go like powers of $e^{i S}$.
\end{enumerate}
Notice that (\ref{TopCorrelator}) is a correlator of the twisted
internal $\N=2$ SCFT. Depending on whether we consider the A-model or
B-model, we have in particular
\begin{align}
&|(\mu_k\cdot G^-)|^2=(\mu_k\cdot G^-)(\bar{\mu}_k\cdot\tilde{G}^+)&&\text{A-model}\nonumber\\
&|(\mu_k\cdot G^-)|^2=(\mu_k\cdot G^-)(\bar{\mu}_k\cdot\tilde{G}^-)&&\text{B-model}\nonumber
\end{align}

Near the conifold singularity it is believed that the topological
string theory is related to compact $c=1$ matter at self-dual radius
coupled to two-dimensional gravity. Indeed it is known that the free
energy of the latter reproduces the leading terms of the topological partition function $F_g$  near
the conifold singularity with the parameter which describes the
deformation of the conifold being related to the Legendre transform of
the cosmological constant of the two-dimensional gravity. It is then natural to
assume that our generalized topological quantities $F_{g,n}$ are
related to some deformation of the $c=1$ system coupled to
two-dimensional gravity. In the heterotic theory, $F_{g,n}$ have been computed in \cite{Morales:1996bp}
when all the matter gauge fields are taken from the dilaton multiplet
(in the dual type II theory this corresponds to the case (2) discussed
above) and it has been shown that near the singularity corresponding
to $SU(2)$ enhancement the leading terms are given by the radius
deformation of the $c=1$ system coupled to two-dimensional gravity away from
the self-dual radius. For completeness we will give a brief review
of \cite{Morales:1996bp} in the following section.

\section{The Amplitude in Heterotic String Theory}\label{Sect:HetAmp}
We now wish to dualize the results of the previous section to
heterotic string theory compactified on $K3\times T^2$. Applying
the same duality mapping as in \cite{Antoniadis:1995zn}, $F_{g,n}$
for arbitrary $g$ and $n$ are expected to start receiving
contributions at the one-loop level. Therefore, in the heterotic
weak-coupling regime which we have already considered in the
previous section, the amplitude
\begin{align}
F_{g,n}=\langle V_{(R)}(p_1)V_{(R)}(\bar{p}_2)\prod_{i=1}^{g-1}V_{(T)}
(\bar{p}_1^i)V_{(T)}(\bar{p}_2^i)\prod_{j=1}^nV_{(F)}(p_1^{j})V_{(F)}
(\bar{p}_2^{j})\rangle_{\text{1-loop}}^{\text{het}}\,,\label{HeteroticAmplitude}
\end{align}
is expected to be identical to (\ref{TopCorrelator}). The vertex operators for gravitons, graviphotons and vector
partners of the dilaton can for example be found in
\cite{Antoniadis:1995zn}. Using the same helicity structure
and polarizations as in Section~\ref{Sect:TypeII} we have
explicitly
\begin{center}
\begin{tabular}{ll}
gravitons: & $V_{(R)}(p_1)=:\left(\partial Z^2-ip_1\chi^1\chi^2\right)
\bar{\partial} Z^2e^{ip_1Z^1}:\,,$\\
& $V_{(R)}(\bar{p}_1)=:\left(\partial \bar{Z}^2-i\bar{p}_1
\bar{\chi}^1\bar{\chi}^2\right)\bar{\partial} \bar{Z}^1
e^{i\bar{p}_1\bar{Z}^1}:\,,$\\&\\
graviphotons: & $V_{(T)}(p_1)=:\left(\partial X -i p_1
\chi^1\Psi\right)\bar{\partial}Z^2 e^{i p_1 Z^1}:\,,$\\
& $V_{(T)}(\bar{p}_1)=:\left(\partial X-i\bar{p}_1\bar{\chi}^1
\Psi\right)\bar{\partial}\bar{Z}^2e^{i\bar{p}_1\bar{Z}^1}:\,,$\\
&\\
gauge fields: & $V_{(F)}(\bar{p}_1)=:\left(\partial X -i\bar{p}_1\bar{\chi}^1
\Psi\right)\bar{\partial}Z^2 e^{i\bar{p}_1\bar{Z}^1}:\,,$\\
& $V_{(F)}(p_1)=:\left(\partial X-i p_1 \chi^1
\Psi\right)\bar{\partial}\bar{Z}^2e^{i p_1 Z^1}:\,.$
\end{tabular}
\end{center}
${}$\\[10pt]
These vertices are written in the $(0)$-ghost picture using the
same complex basis for the space-time coordinates $(Z^1,\bar{Z}^1,
Z^2,\bar{Z}^2)$ and their fermionic partners $(\chi^1,\bar{\chi}^1,
\chi^2,\bar{\chi}^2)$ as in Section~\ref{Sect:TypeII}. The (complex)
coordinate on $T^2$ is called $(X,\bar{X})$ with its superpartner
$(\Psi,\bar{\Psi})$.

In order to compute (\ref{HeteroticAmplitude}) it suffices to consider
the odd spin-structure since the even one gives essentially the same
result. In this case both gravitons need to contribute the fermion
bilinear piece in order to soak up the corresponding zero modes of
the space-time fermions. Due to the presence of a Killing spinor on
the world-sheet torus (see \cite{Antoniadis:1995zn}), we pick one
of the graviphotons to be inserted in the $(-1)$ picture
\begin{align}
V_{(T)}^{(-1)}(p_1)=e^{-\phi}\Psi\bar{\partial}Z^2e^{i p_1  Z^1}\,,
\end{align}
which will be accompanied by a picture changing operator. The latter,
in order to soak the $\Psi$ zero-mode needs to contribute $e^{\phi}
\bar{\Psi}\partial X$. Since the remaining $\Psi$ cannot be
contracted, it follows that all gauge-field vectors will contribute
their $\partial X$ piece, which will just give zero modes. Therefore,
the only remaining non-trivial piece is the correlator of the
space-time bosons. Extracting the appropriate piece proportional to
$p_1^2\bar{p}_1^2\prod_{i=1}^{g-1} p_1^i\bar{p}_1^i\prod_{j=1}^n\bar{p}_1^j
p_1^j$ it takes the form
\begin{align}
\mathcal{C}_{g,n}=\langle\prod_{i=1}^g\int d^2x_i Z^1\bar{\partial}
Z^2 (x_i)\,\int d^2y_i \bar{Z}^1\bar{\partial} \bar{Z}^2(y_i)\,
\prod_{j=1}^n \int d^2 u_j\bar{Z}^1\bar{\partial} Z^2 (u_j)\,
\int d^2v_j Z^1\bar{\partial}\bar{Z}^2(v_j)\rangle\,.\nonumber
\end{align}
Since this is just a free field correlator, it can be computed by
evaluating the generating functional
\begin{align}
&G(\epsilon_-,\epsilon_+)=\sum_{g=1}^\infty\sum_{n=1}^\infty\frac{\epsilon_-^{2g}\epsilon_+^{2n}
\mathcal{C}_{g,n}}{(g!n!)^2\tau_2^{2g+2n}}\,=\nonumber\\
&=\frac{\int \prod_{i=1}^2\mathcal{D}Z^i\mathcal{D}\bar{Z}^i
  \text{Exp}
\left(-S+\frac{\epsilon_-}{\tau_2}\int (Z^1\bar{\partial}Z^2+\bar{Z}^2
\bar{\partial}\bar{Z}^1)+\frac{\epsilon_+}{\tau_2}\int
(\bar{Z}^1\bar{\partial}Z^2+\bar{Z}^2\bar{\partial}Z^1)\right)}
{\int \prod_{i=1}^2\mathcal{D}Z^i\mathcal{D}\bar{Z}^i \text{Exp}
\left(-S\right)}\,,\label{FunctionalIntegral}
\end{align}
where $S$ is the usual free-field action $S=\sum_{i=1,2}\frac{1}
{\pi}\int d^2x(\partial Z^i\bar{\partial}\bar{Z}^i+\partial
\bar{Z}^i\bar{\partial}Z^i)$ and $\epsilon_\pm$ are two
arbitrary parameters. Notice that all terms in
(\ref{FunctionalIntegral}) are bilinear in the fields. Thereby
the functional integral is Gaussian and can be evaluated
explicitly. This is essentially done in the same way as in
\cite{Morales:1996bp,Antoniadis:1995zn} and we can therefore
immediately state the result
\begin{align}
G(\epsilon_-,\epsilon_+)=&\frac{2\pi i(\epsilon_-+\epsilon_+)\bar{\eta}^3}
{\bar{\vartheta}_1(\epsilon_-+\epsilon_+,\bar{\tau})}\cdot
\frac{2\pi i(\epsilon_--\epsilon_+)\bar{\eta}^3}{\bar{\vartheta}_1
(\epsilon_--\epsilon_+,\bar{\tau})}\,e^{-\frac{\pi}{\tau_2}(\epsilon_+^2+\epsilon_-^2)}\,.
\end{align}
Upon introducing also a generating functional for the amplitudes
\begin{align}
F(\epsilon_-,\epsilon_+)=\sum_{g=1}^\infty\sum_{n=1}^\infty \epsilon_-^{2g}\epsilon_+^{2n}F_{g,n}\,,
\end{align}
we can state the final result
\begin{align}
F(\epsilon_-,\epsilon_+)\!=\!\!\!\int \frac{d^2\tau}{\tau_2}\frac{\bar{E}_4\bar{E}_6}
{\bar{\eta}^{24}}\sum_{\Gamma^{(2,2)}}\left(\frac{2\pi i(\epsilon_-+\epsilon_+)
\bar{\eta}^3}{\bar{\vartheta}_1(\tilde{\epsilon}_- +\tilde{\epsilon}_+,
\bar{\tau})}\right)\left(\frac{2\pi i(\epsilon_--\epsilon_+)
\bar{\eta}^3}{\bar{\vartheta}_1(\tilde{\epsilon}_- -\tilde{\epsilon}_+,
\bar{\tau})}\right) e^{-\frac{\pi}{\tau_2}(\tilde{\epsilon}_-^2+\tilde{\epsilon}_+^2)}
q^{\frac{1}{2}|P^L|^2}\bar{q}^{\frac{1}{2}|P^R|^2}\label{HetOneLoop}
\end{align}
where we have introduced ~$\tilde{\epsilon}_{\pm}
=\epsilon_{\pm}\tau_2  P^L/\sqrt{(T-\bar{T})(U-\bar{U})}$~ 
with $P^L$ and $P^R$ the lattice momenta
of the internal $\Gamma^{(2,2)}$ lattice, while $q=e^{2\pi i\tau}$. Notice that we have
written this amplitude at a point in moduli space with no Wilson
lines in the right moving sector being switched on. $P_L$ is the
central charge of the $\N=2$ algebra and is given by
\begin{equation}
P^L= \frac{1}{\sqrt{(T-\bar{T})(U-\bar{U})}} (n_1+n_2 \bar{T}\bar{U}+m_1
\bar{T}+m_2 \bar{U})\,,
\end{equation}
where $m_1,m_2,n_1,n_2$ are integers and $T$ and $U$ are the K\"ahler
and complex structure moduli of the compactification torus.

Near an $SU(2)$ enhanced symmetric point in moduli space $P_L \to 0$
for two lattice points in $\Gamma^{(2,2)}$ ($n_1=n_2=0,\,m_1=-m_2=\pm1$). In the vicinity of these points the leading singularity
for $F_{g,n}$ can be easily computed from (\ref{HetOneLoop})
as $\epsilon_-^{2g} \epsilon_+^{2n}$ term in the expansion of
\begin{equation}
F\sim    \int\frac{ d\tau_2}{\tau_2} \frac{\pi
  \epsilon_1}{\sin(\pi \epsilon_1 \tau_2)}\frac{\pi
  \epsilon_2}{\sin(\pi \epsilon_2  \tau_2)} e^{-\tau_2 \mu}\,,
\label{singularity}
\end{equation}
where $\mu= \sqrt{\frac{i}{\pi}}\sqrt{(T-\bar{T})(U-\bar{U})}
\bar{P}^L\sim \sqrt{\frac{i}{\pi}}(T-U)$ and $\epsilon_1=\epsilon_++\epsilon_-$ and
$\epsilon_2=\epsilon_+-\epsilon_-$ and we have also rescaled ~$\tau_2
\bar{\mu} \rightarrow \tau_2$. The 
leading singularity as $\mu \rightarrow 0$ in the
$\epsilon_-^{2g} \epsilon_+^{2n}$ term is given by
$\mu^{2-2g-2n}$. Note that this is holomorphic in the relevant vector modulus
$T-U$ and it is related to the fact that our vertices for the
graviphoton and dilaton field strengths involve only $\partial X$ in
the $T^2$ direction which  contributes $P_L$ in the correlation
function. 
This expression is just the free energy of the $c=1$ system
coupled to 2-dimensional gravity with the cosmological constant $\mu$ and at radius
$|\epsilon_1/\epsilon_2|$  in units where the self dual radius is
1~\cite{c1}. Note that for $\epsilon_+=0$, which reduces to the standard $F_g$,
we have $\epsilon_2=-\epsilon_1$ giving rise to the free energy of the
$c=1$ system at the self-dual radius. Equation (\ref{singularity})
exhibits certain symmetries generated by $(\epsilon_1,\epsilon_2)
\rightarrow (-\epsilon_1,\epsilon_2)$
and $(\epsilon_1,\epsilon_2) \rightarrow (\epsilon_2,\epsilon_1)$.\footnote{The resulting
  symmetry group is a discreet non-abelian group $Dihedral_4$ of 8 elements $(\pm
  1, \pm \sigma_1, \pm \sigma_3, \pm i \sigma_2)$.} In particular this
implies that (\ref{singularity}) is invariant under the exchange of
$\epsilon_+$ and $\epsilon_-$ and  is even in $\epsilon_1$, $\epsilon_2$
as well as $\epsilon_+$ and $\epsilon_-$. The latter is to be expected
since $\epsilon_+$ and $\epsilon_-$ couple to self-dual and anti-self
dual field strengths and a Lorentz invariant string effective
action must contain even numbers of self-dual and anti-selfdual 
tensors.

This expression is very similar to the perturbative part of the free
energy of the pure $\N=2$ $SU(2)$ Yang-Mills theory in the
$\Omega$-background given in (A.7) of \cite{Nekrasov:2003rj} with the
identification $2\pi i \epsilon_{1,2} \rightarrow
\epsilon_{1,2}$. There instead of the $\sin$ factors in the denominator
of (\ref{singularity}) they have the factor
\begin{equation}
F\sim    \int\frac{ d\tau_2}{\tau_2} \frac{\pi
\epsilon_1}{\sin(\pi \epsilon_1 \tau_2)}\frac{\pi
\epsilon_2}{\sin(\pi \epsilon_2 \tau_2)} e^{-\tau_2 \mu} e^{-i\pi (\epsilon_1+\epsilon_2)\tau_2}    
\label{extraphase}
\end{equation}
This expression has only the symmetry  $\epsilon_1 \leftrightarrow
\epsilon_2$. It is even in $\epsilon_-$ but not in $\epsilon_+$ due to
the extra phase $e^{-i\pi (\epsilon_1+\epsilon_2)\tau_2}$. 

\section{Concluding remarks}\label{Sect:Conclusions}

In this paper we have analyzed the $\N=2$~ F-terms of the type
$W^{2g} \Upsilon^n$ where $W$ is the chiral Weyl superfield and
$\Upsilon$  are chiral projectors of  real functions of vector
multiplets. These terms give rise to couplings of the form
$(R^-)^2(T^-)^{2g-2} (F^+)^{2n}$ with $R$, $T$ and $F$ being the
Riemann tensor, graviphoton and matter vector field strengths. 
In the special case where the vector multiplets only
involve the heterotic dilaton $S$, these amplitudes have been computed at one
loop in heterotic string
\cite{Morales:1996bp} where it was shown that in the field theory limit
they are given by the radius deformation of $c=1$ matter coupled to
2-dimensional gravity.  This result is perturbative in the
heterotic dilaton and should receive non-perturbative corrections.
In the present work we have constructed these
amplitudes in type II theory for general $\Upsilon$ involving
arbitrary vector multiplets, where they appear at genus $g$ 
and are given by certain correlation functions in the twisted
topological theory which is essentially the square of the supersymmetric
sector of the ${\cal{N}}=1$
heterotic topological amplitude constructed earlier in \cite{Antoniadis:1996qg}. Due to
the fact that the type II dilaton is a hypermultiplet we expect this genus
$g$ result to be exact.  

Remarkably, the field theory limit in the heterotic side (when the
matter vector fields are taken from dilaton multiplet) reproduces
the perturbative part of
Nekrasov results for the ${\cal{N}}=2$ gauge theories in the general
$\Omega$ background 
involving 2 complex parameters $\epsilon_{\pm}$, up to a phase depending
on $\epsilon_+$, which
might be attributable to different regularization schemes.  In string theory $\epsilon_-$
couples to the graviphoton $T^-$ and $\epsilon_+$ to
$F_S^+$. It might appear strange that while $\epsilon_-$
couples to the anti-self dual graviphoton, $\epsilon_+$ couples to
the self-dual dilaton vector. However in the heterotic theory this is
quite natural for two reasons:
\begin{enumerate}
\item The left moving ({\em i.e.} in the supersymmetric sector) the
vertices of the anti-self dual graviphoton and the self-dual dilaton
vector are identical and in fact both of them contribute $P_L$ lattice
momentum. The difference in the two vertices appears in the bosonic
sector where they give anti-self dual and self dual Lorentz currents 
respectively. Thus setting $\epsilon_+=\pm \epsilon_-$
({\em i.e.} $\epsilon_1$ or $\epsilon_2$ equal to zero) we find a manifest
$SO(2)$ invariance as happens in the gauge theory.
\item More importantly,
since the anti-selfdual 
graviphoton and selfdual dilaton vector vertices contribute $P_L$, this amplitude
carries $(2g+2n-2)$ $P_L$. Therefore, as  discussed in the last
section, near  the $SU(2)$ enhancement $\mu \rightarrow 0$, where $\mu$
is the holomorphic vector modulus, the leading singularity has the
behaviour $\mu^{2-2g-2n}$ exactly as in the gauge theory. For example,
if some of the vertices had been that of selfdual graviphoton, they
would have contributed $\bar{P}_L$, and this  would
have given rise to factors of  $1/\bar{\mu}$ to the singularity. The
answer then would not be holomorphic in vector moduli 
as opposed to
the gauge theory result of \cite{Nekrasov:2002qd,Nekrasov:2003rj} as
expected from the fact that this result gives the prepotential of the gauge theory.
\end{enumerate}
This generalizes  the
case $\epsilon_+=0$ for which it is known that $F_{g,0}$, in the field
theory limit and weak heterotic coupling, reproduces the
perturbative part of Nekrasov's result. For $\epsilon_+ \neq 0$, while
the heterotic string (as well as the type II) result is even in both $\epsilon_+$ and $\epsilon_-$ (this is
a consequence of the Lorentz invariance of the string effective
action), Nekrasov's
result is even only in $\epsilon_-$ and due to the extra phase in 
(\ref{extraphase}) it contains all powers in $\epsilon_+$. On the other hand, 
the gauge theory action given in eqs.~(2.1) and (2.3) of
\cite{Nekrasov:2003rj} is Lorentz invariant provided that
transformations of the ${\cal{N}}=2$ gauge theory fields are compensated
by a simultaneous Lorentz transformation of the $\Omega$ background.
One would therefore expect that the quantum corrections to the
prepotential must also have this symmetry. Since $\epsilon_{\pm}$
transform under the~ $SU(2)_L \times SU(2)_R$~ of the 
Lorentz group, the invariants must be quadratic in $\epsilon_+$ as
well as $\epsilon_-$. The appearance of odd powers in $\epsilon_+$ in
the gauge theory result is therefore surprising and might be a
consequence of the regularization procedure. In fact the generalized
Gopakumar-Vafa formula (eq.(2.6) of \cite{Iqbal:2007ii} ) is based on
the one loop Schwinger result of a particle
in representation ~$\cal R$~ of ~$SU(2)_L \times SU(2)_R$~ 
in the background of constant gauge field strength
$F=F_1 dx^1 \wedge dx^2 + F_2 dx^3 \wedge dx^4$:
\begin{equation}
\int \frac{ds}{s} \frac{Tr_{\cal R} (-1)^{\sigma_L+\sigma_R} e^{-s
    m^2} e^{-2se(\sigma_L F_++ \sigma_R F_-)}}{(2 \sinh(se F_1/2))(-2 \sinh(seF_2/2)}\,,
\label{GV}
\end{equation}
where $F_{\pm}$ are the selfdual and anti-selfdual parts of $F$, $\sigma_{L,R}$ are the weights of the $SU(2)_{L,R}$ in the
representation $\cal R$ and $m$ and $e$ are the particle's mass and charge respectively. In  each representation ~${\cal R }= (j_L,j_R)$~ the above
expression is even in $F_+$ and $F_-$ since in the trace, $\sigma_{L,R}=-j_{L,R},\ldots, j_{L,R}$. This is consistent with the
result of the heterotic string  as $ F_{\pm}$ in (\ref{GV}) are
related to the background 
$ \epsilon_{\pm}$.

The type II result we have obtained in this paper, when applied for the
vector multiplets corresponding to the dual heterotic dilaton $S$, also
captures the non-perturbative contributions that go like exponentials
of $S$. As pointed out in Section 3, in order to extract the 1PI
effective action term, one needs to subtract the reducible graph shown
in fig.1. Note that unlike the
heterotic case, the type II result is not symmetric under the exchange of
$\epsilon_+$ and $\epsilon_-$. Indeed, while the power of $\epsilon_-$ counts 
twice the genus $g$ of the Riemann surface,~$\epsilon_+$ counts the
number of vertices $2n$ in the twisted theory. This is also true in
the gauge theory result (at the non-perturbative level). 
It will be  interesting to see how the  type II amplitudes
considered here are related to the generalized Gopakumar- Vafa formula
\cite{Iqbal:2007ii} and to non-perturbative part in the gauge theory
prepotential \cite{Nekrasov:2002qd,Nekrasov:2003rj} in the presence of arbitrary $\epsilon_+$ and $\epsilon_-$.

In the heterotic side, near the enhanced symmetric point, the leading
singularity is given by the free energy of the radius deformed (away
from the self-dual radius) $c=1$ CFT coupled to 2-dimensional
gravity. By heterotic - type II duality, the type II amplitudes we
have considered, when specialized to all the matter vector fields to
be the  heterotic dilaton vector fields, near the conifold should also 
describe the radius deformation of the above system. While in the type
II it is expected
that turning on $\epsilon_+$ should be some deformation of the
self-dual $c=1$ system coupled to 2-dimensional gravity, it will be
interesting to establish  directly in  type II theory ({\em i.e.} without
resorting to heterotic duality) that the deformation is given by the
change of radius.

\section*{Acknowledgements}
We would like to thank S.~Shatashvili for enlightening discussions and N.~Nekrasov for helpful correspondence. I.A. acknowledges also discussions with A.~Gerasimov.
Work supported in part by the European Commission under the ERC Advanced Grant 226371 and the contract PITN-GA-2009-237920 and in part by the CNRS grant GRC APIC PICS 3747. The research of S.H. has been supported by the Swiss National Science Foundation. T.R.T.\ is grateful to CERN Theory Unit for financial support and kind hospitality during the first part of his sabbatical leave. His research is partially supported
by the U.S. NSF Grants PHY-0600304, PHY-0757959. Any opinions, findings,
and conclusions or recommendations expressed in this material are
those of the authors and do not necessarily reflect the views of the U.S. National Science Foundation.


\end{document}